\documentclass[superscriptaddress,twocolumn,prl]{revtex4-1}
\usepackage{graphicx}                   
\usepackage{hyperref}                   
\usepackage{amsmath,amssymb}            
\usepackage{multirow}	
\usepackage{xcolor}	
\usepackage{amsmath,color}

\graphicspath{{s/}}					

\DeclareMathAlphabet{\mathpzc}{OT1}{pzc}{m}{it}

\begin{document}

	\author{Abbas Ali Saberi}
	\email{ ab.saberi@ut.ac.ir \& saberi@pks.mpg.de}
	\affiliation{Department of Physics, University of Tehran, P. O. Box 14395-547, Tehran, Iran}
	\affiliation{Max Planck Institute for the Physics of Complex Systems, 01187 Dresden, Germany}
 
 \author{Sina Saber}
	\affiliation{Department of Physics, University of Tehran, P. O. Box 14395-547, Tehran, Iran}
	
	\author{Roderich Moessner}
	\affiliation{Max Planck Institute for the Physics of Complex Systems, 01187 Dresden, Germany}
	
	\title{Interaction-correlated random matrices}
	
	\date{\today}
	
	\begin{abstract}
		We introduce a family of random matrices where correlations between matrix elements are induced via interaction-derived Boltzmann factors. Varying these yields access to different ensembles. 
		We find a universal scaling behavior of the finite-size statistics characterized by a heavy-tailed eigenvalue distribution whose extremes are governed by the Fréchet extreme value distribution for the case corresponding to a ferromagnetic Ising transition.
		The introduction of a finite density of nonlocal interactions restores standard random matrix behavior. Suitably rescaled average extremes,  playing a physical role as an order parameter, can thus discriminate aspects of the interaction structure; they also yield further nonuniversal information. In particular, the link between maximum eigenvalues and order parameters offers a potential route to resolving long-standing problems in statistical physics, such as deriving the exact magnetization scaling function in the 2D Ising model at criticality.
	\end{abstract}
	\maketitle

 \textit{Introduction.}---
Random matrix theory (RMT) \cite{akemann2011oxford} has played a significant role as a model for the statistical properties of a wide range of complex systems in diverse topics including nuclear and theoretical physics \cite{wigner1951statistical}, financial mathematics \cite{laloux2000random}, neuroscience \cite{wainrib2013topological}, telecommunications \cite{couillet2011random}, price fluctuations in the stock market \cite{plerou1999universal, laloux1999noise}, EEG data of the brain \cite{vseba2003random}, variation of different atmospheric parameters \cite{santhanam2001statistics}, ecological communities \cite{baron2023breakdown}, and complex networks \cite{jalan2007random}. The success of RMT in describing complex systems is based on one of the ideas that the collective behavior of many uncorrelated random variables and their extremes exhibits universal patterns. However, in most of the physical systems mentioned above, the underlying random variables describing the microscopic constituents are typically subject to short- or long-range correlations and interactions. Indeed, discovering the interaction patterns underpinning huge amounts of data collected from complex systems in biology, ecology, sociology, economy, or indeed experimental quantum measurements, is a central problem in artificial intelligence and inverse statistical problems, with the aim of modeling real-world systems \cite{national2013frontiers, janiesch2021machine, mahmud2021deep}. This poses a major challenge for reconstructing the correlation structures from the emerging collective behavior of individual elements, where traditional data analysis tools struggle to extract underlying features.

An intriguing question is whether RMT can reveal or characterize underlying interactions and correlations in a physical system. For this, it is a first step necessary to consider a scheme of random matrices with correlated rather than fully independent elements.
	While the RMT of uncorrelated random variables is well understood, less is known about the universal properties of the more complex random matrices with long-range correlated elements.
	In the past, cases of random matrices in which the entries are allowed to exhibit a certain correlation structure have been studied. Among them are Euclidean matrix ensembles possessing triangular-like correlations \cite{mezard1999spectra}, random matrices with general slowly decaying correlations \cite{erdHos2019random, ajanki2019stability}, ensembles from nonextensive entropy \cite{toscano2004random}, as well as almost uncorrelated ensembles \cite{hochstattler2016semicircle}, etc. In particular, random matrix ensembles with slowly decaying correlations (polynomial decay) are studied in \cite{erdHos2019random, ajanki2019stability} using a novel cumulant expansion method. These works prove local laws and bulk universality, demonstrating that eigenvalue statistics remain universal and independent of the specific correlation structure, whether correlations decay rapidly (exponentially) \cite{ajanki2019stability} or slowly (polynomially) \cite{erdHos2019random}.

Here, we introduce what we call interaction-correlated ensembles of random matrices, where the correlations between matrix elements are derived from an underlying physical system governed by the Boltzmann measure of a microstate in a two-dimensional many-body system at thermal equilibrium. This probability measure is defined by the system's energy function and temperature, allowing us to control the nature of the interactions that induce correlations in the matrix ensembles. Specifically, we use the two-dimensional (2D) Ising model at temperature $T$ as the foundation for generating these ensembles. The matrix ensembles, therefore, serve as a mathematical abstraction of the underlying interacting many-body system, capturing a wide spectrum of correlation behaviors. This approach provides a controlled framework for studying how the intrinsic interactions in the Ising model manifest in the statistical properties of the eigenvalue spectra of the corresponding random matrices.

It is important to note that our approach is distinct from traditional random matrix models which represent the Hamiltonians of quantum interacting many-body systems in Hilbert space, and which focus on specific matrix structures arising from few-body interactions. Unlike those models, our construction leverages the equilibrium properties of classical systems to model real-space correlations in a random matrix ensemble, offering a novel perspective on how universal spectral features in RMT, particularly near criticality, are influenced by the presence of nontrivial long-range correlations.

The main results of this letter, based on extensive simulations and supported by analytical arguments, are as follows:
(1) An ensemble of random matrices derived from Boltzmann weights at criticality asymptotically produces an emergent bell-shaped bulk spectrum with a universal heavy tail, contrasting with the semicircle-law observed in the off-critical regime.
(2) At criticality, the extreme eigenvalue statistics converge to a universal Fréchet distribution, deviating from the typical Tracy-Widom distribution in traditional RMT.
(3) The spectrum of the mean-field limit of the interacting system at criticality aligns with the predictions of standard RMT.
(4) Our numerical results demonstrate that the rescaled average extreme eigenvalue serves as an order parameter, capturing both universal and non-universal aspects of the system's interaction structure.

	\textit{Model.}---
We explore the spectral properties of a random-matrix ensemble where the matrix elements correspond directly to the spins of a 2D ferromagnetic Ising model, without an external field, on a square lattice of linear size $L$ in thermal equilibrium at temperature $T$. 
Each element $\mathcal{M}'_{ij}$ of the real random matrix $\mathcal{M}'$ takes on spin values $s=\pm1$ at the lattice site $(i,j)$, where $1\le i,j\le L$. Periodic boundary conditions are used. We perform a finite-size scaling analysis to study the eigenvalue spectrum of an ensemble of such matrices after symmetrization $\mathcal{M}=(\mathcal{M}'+\mathcal{M}'^T)/2$ to ensure real eigenvalues. Thus, the real symmetric matrix $\mathcal{M}$ has $\frac{1}{2}L(L+1)$ independent elements $\mathcal{M}_{ij}$ which can take values $\pm1$ or $0$, while the diagonal elements $\mathcal{M}_{ii}=\pm1$. The zero-mean condition only holds for $T\ge T_c$ where the average magnetization $\textsf{m}(T)=\langle s\rangle$ vanishes for $L\rightarrow\infty$.
	Correlations between spins at distance $r$ are captured by the spatial correlation function $G(r)$. In the off-critical regimes, the correlations are short-range with an exponentially decaying form $G(r) \approx r^{-\tau}\exp(-\frac{r}{\xi})$, with $\tau=1/2$ for $T>T_c$ and $\tau=2$ for $T<T_c$ \cite{henkel1999conformal}. At the critical point, $T=T_c$  the correlation length diverges,  $\xi\rightarrow\infty$, and $G(r)\approx r^{-d+2-\eta}$ has a power-law form with the correlation length exponent $\eta=1/4$. We also consider the 2D Ising model with nonlocal interactions, which leads to a mean-field limit with $\eta=0$, and investigate the corresponding random matrix ensembles.
 
As we will demonstrate, Ising interactions provide a natural avenue towards the simplest correlated random matrices with qualitatively distinct behavior compared to conventional random matrix ensembles, while minimizing unnecessary complexity. This approach facilitates addressing the overarching question of how correlations impact the notion of universality within a broader class of correlated matrices. Our study is broadly related, for instance,  to disordered spin systems and Anderson localization, where random matrices with correlated elements emerge from long-range physical processes. The correlations in the matrix elements then reflect spatial correlations in the distribution of the disordered spin couplings or hopping matrix elements. In the case of a neural network, such correlations will be related to its functionality, and can generally be very complex.
	
	\begin{figure}[t]
		\centering
		\includegraphics[width=3.4in]{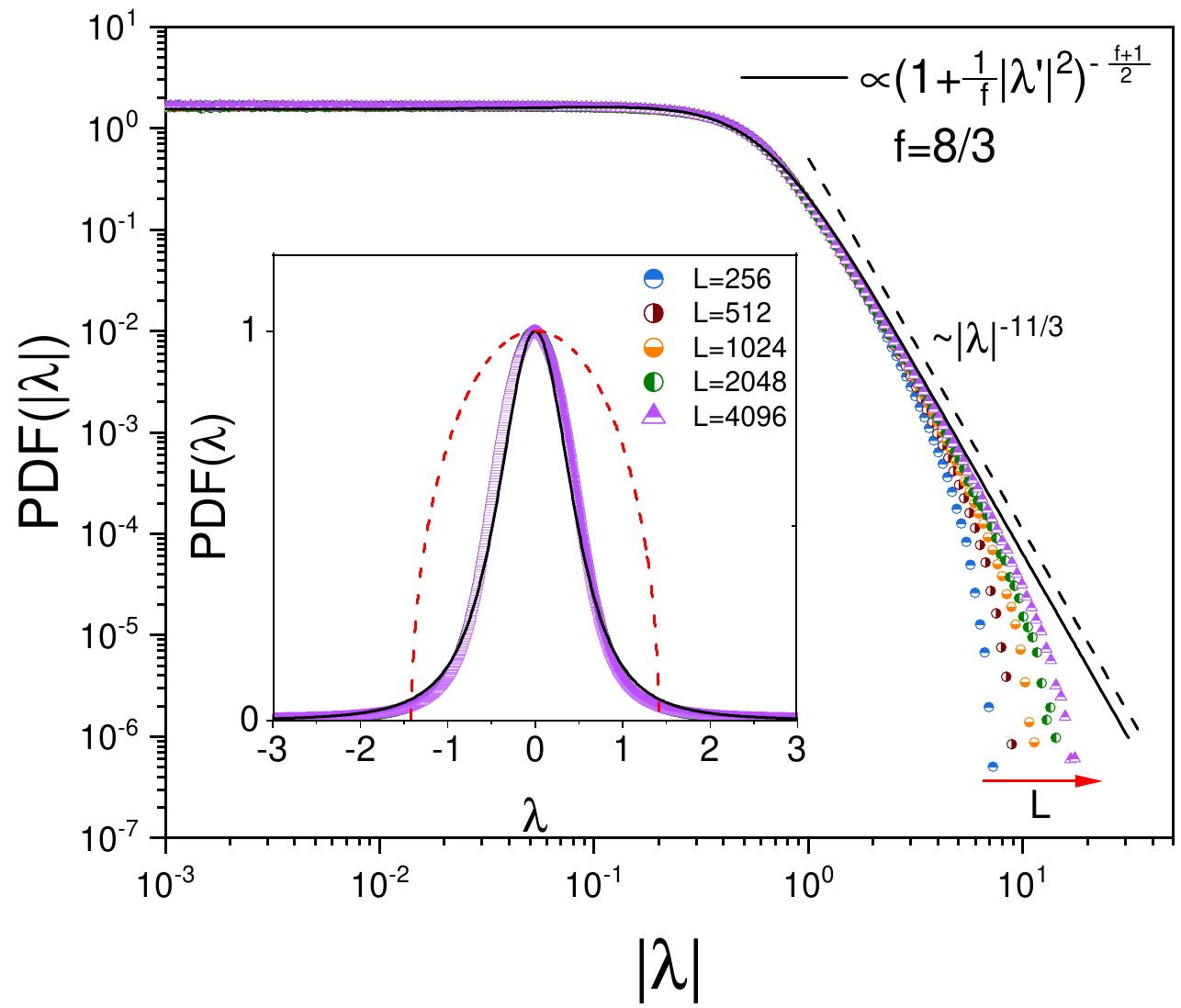}
		\caption{Main: Distribution of the absolute eigenvalues of an ensemble of rescaled (see text) symmetric random matrices $\{\mathcal{M}\}$ as a realization of a 2d Ising model on a square lattice of linear size $L=2^8, 2^9, 2^{10}, 2^{11}$ and $2^{12}$, at the critical point $T=T_c$. The ensemble size is $10^5$ for the smallest and $10^4$ for the largest system size. The plots collapse for all system sizes. The distribution has a heavy tail $\sim |\lambda|^{-\text{f}-1}$ with tail exponent $11/3$ (dashed line) in the large size limit. The solid line shows our theoretical prediction for the asymptotic probability distribution compatible with a $t$-distribution with $\text{f}=8/3$ degrees of freedom (Eq. (\ref{t-dist})).  Inset: Distribution of the eigenvalues for different system sizes compared with the corresponding $t$-distribution Eq. (\ref{t-dist}) (solid line). The red dashed line shows the semi-circle law of standard RMT for the PDF of bulk eigenvalues.}
		\label{fig1}
	\end{figure}

	\begin{figure}[t]
		\centering
		\includegraphics[width=3.4in]{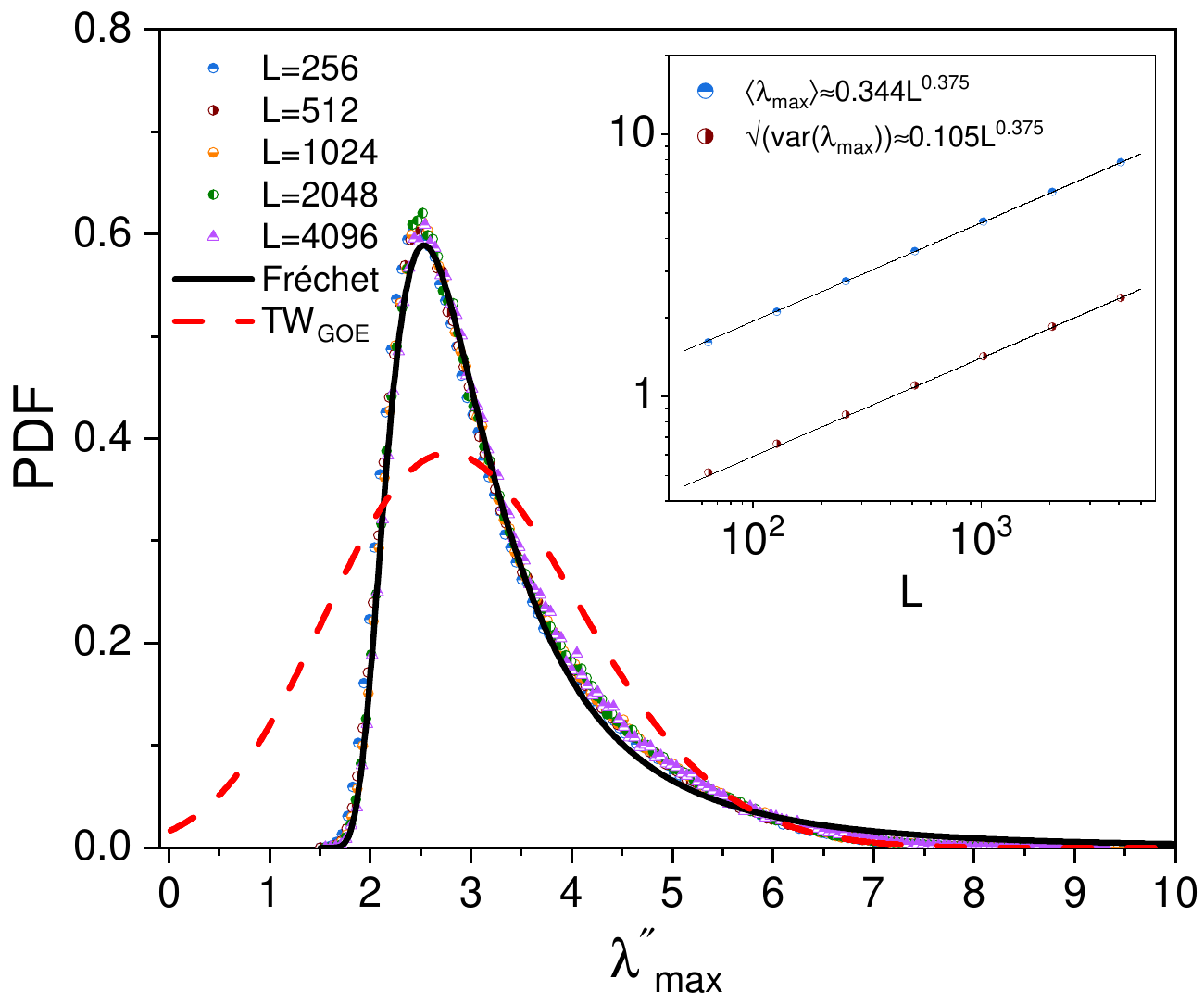}
		\caption{Finite-size scaling collapse of the maximum eigenvalue $\lambda_{\text{max}}$ distributions for an ensemble of rescaled symmetric random matrices $\{\mathcal{M}\}$ at $T=T_c$. For each system size $L$, we have generated $10^6$ independent samples. The scaling exponent $\text{b}=0.375(15)$ is measured by examining the scaling relation $\langle \lambda_{\text{max}}\rangle\sim L^{\text{b}}$ shown in the inset. This exponent is further verified by examining the scaling relation of the standard deviation of $\lambda_{\text{max}}$ with system size (Inset). The solid line in the main panel shows our theoretical predictions based on Eqs. (\ref{univ_Frechet}) and (\ref{Frechet}) for the universal function describing the collapsed data as a Fréchet extreme value distribution with shape parameter $\text{k}=1/\text{f}=3/8$. The red dashed line shows the prediction by the standard RMT for comparison.}
		\label{fig2}
	\end{figure}

	\textit{Results.}---We have produced an ensemble of ($10^4$ for the largest size $L=2^{12}$ to $10^5$ for the smallest size $L=2^8$) thermally equilibrated samples of spin configurations at temperature $T$, and we study the eigenvalues of the abovementioned  random matrix $\{\mathcal{M}\}$. To be comparable with standard random matrix theory, the matrices are rescaled by a factor of $1/\sqrt L$ so that the edges of the semicircle distribution of the bulk eigenvalues lie in the interval $[-\sqrt 2, \sqrt 2]$.

	\textit{Critical ensemble.}
	We plot the probability distribution function (PDF) of the eigenvalues at $T=T_c$ for different sizes from $L=2^8$ to $2^{12}$ in Figure \ref{fig1}. All data for different sizes collapse. The PDF of eigenvalues possesses the following characteristic and remarkable features: (i) It is bell-shaped and symmetric about zero with zero skewness, like the Gaussian distribution, but (ii) it has heavier tails prone to producing more extreme events. As shown in the main Figure \ref{fig1} in a double logarithmic scale, the PDF of the absolute eigenvalues has a power-law tail $\sim |\lambda|^{-\text{f}-1}$ with the tail exponent $11/3$ based on our theoretical arguments that will be provided in the following. The deviation at the tails is a finite-size effect, with the data approaching the asymptotic power law as size increases. These are the key properties of the so-called $t-$distribution which belongs to a family of probability distributions distinguished by the exponent $\text{f}$, \begin{equation}\label{t-dist}
	\mathcal{P}(\lambda)=a_0\Big(1+\frac{1}{\text{f}}\lambda'^2\Big)^{-\frac{\text{f}+1}{2}},
	\end{equation}
	with $a_0$ being the normalizing constant, and the prime denotes the rescaling of the eigenvalues by a factor related to the variance $\sigma^2$ of $\lambda$. Our finite-size analysis for  $\sigma^2$ finds that it converges to $\sim1/2$ as $L\rightarrow\infty$ by the scaling relation $(1/2-\sigma^2)\sim L^{-\theta}$ with $\theta\simeq 0.25$ (see the upper panel in the Supplementary Figure S1). This is in perfect agreement with the exact correlation exponent $\eta=1/4$ for the 2D Ising model. Therefore, there will be no free parameters in our proposed probability density function (\ref{t-dist}). $\mathcal{P}(\lambda)$ in Eq. (1) is even, so the distribution of the absolute eigenvalues $\mathcal{P}(|\lambda|)$ is 2 times $\mathcal{P}(\lambda)$, supported on $[0, \infty]$. To illustrate the match of our proposed functional form, we also display the density function (\ref{t-dist}) with $\text{f}=8/3$. To contrast our finding with standard RMT, we have also plotted the semicircle law.

	For the standard semicircle law, one expects the typical fluctuations of the maximum eigenvalues $\lambda_{\text{max}}$ to be given by the $\mathcal{TW}_{1}$ distribution \cite{tracy1994level} over a region of width $\sim\mathcal{O}(L^{-2/3})$ around the upper edge $\sqrt{2}$ of the Wigner sea. This means that for $L\rightarrow\infty$, the average maximum eigenvalue $\langle\lambda_{\text{max}}\rangle=\sqrt{2}$. However, our result (\ref{t-dist}) shows that the long-range correlations between the matrix elements at the critical point $T=T_c$ drive the spectral density to a new universal fixed point, characterized by an emergent heavy-tailed spectrum that allows for the occurrence of very large extreme eigenvalues. (our analysis shows that the excess kurtosis of the bulk eigenvalues diverges with system size $\propto\sqrt{L}$ at $T_c$---see the lower panel in the Supplementary Figure S1). 
	
	We find that the average maximum eigenvalue shows a power-law relation with system size as $\langle\lambda_{\text{max}}\rangle=aL^b$, with $a=0.344(20)$ and exponent $b=0.375(15)$ (inset of Fig. \ref{fig2}).
	We use this scaling relation in the main Figure \ref{fig2} to show the finite-size collapse of the largest eigenvalue distributions computed for system sizes  $L=2^8$ to $2^{12}$. This suggests the following finite-size scaling form for the distribution of the largest eigenvalue at $T_c$ \begin{equation}\label{univ_Frechet}
	\mathcal{P}(\lambda_{\text{max}})=L^{-b}\mathcal{F}\Big(\lambda_{\text{max}}L^{-b}\Big),
	\end{equation}
	where $\mathcal{F}(\cdot)$ is a universal scaling function, shown in the following to be the Fréchet extreme value distribution. 
 	This contrasts with conventional RMT with its semicircle bulk and corresponding GOE edge spectrum (Fig.~\ref{fig2}).

	\textit{Analytical arguments.}
Here, we utilize the developed tools within the context of extreme value theory \cite{majumdar2020extreme} to determine the only free parameter 'f' in Eq. (\ref{t-dist}).
 Considering that the scaled average extreme eigenvalue acts as an order parameter, an analytical framework can be developed to describe the spectral properties of the Ising model at $T_c$. For $\lambda\gg 1$, Eq.  (\ref{t-dist}) predicts a heavy-tailed distribution for the largest eigenvalues $\mathcal{P}(\lambda)\sim |\lambda|^{-\text{f}-1}$. These atypical eigenvalues are collected from an ensemble of independent samples; however, despite the eigenvalue correlations within a single realization, the power-law tail observed in the eigenvalue spacing distribution (in contrast to the Wigner-Dyson distribution in traditional RMT) suggests that the largest eigenvalues are more sparsely spaced. This, along with the scaling behavior of the extreme eigenvalues, provides strong evidence for the applicability of extreme value theory (EVT) in our model, specifically aligning with the Fréchet universality class. This result is consistent with prior analytical demonstrations in  correlated systems, such as in \cite{evans2008condensation}, where a Fréchet distribution emerges at the critical point despite global mass conservation constraints.\\ Now let $Q_L(\lambda)=\text{Prob}\big[\lambda_\text{max}<\lambda\big]$ be the cumulative distribution function of $\lambda_\text{max}$. If $\lambda_\text{max}<\lambda$, then all other eigenvalues should necessarily be less than $\lambda$, and since they are independent,  $Q_L(\lambda)=\big[1- \int_{\lambda}^{\infty} \mathcal{P}(\lambda)d\lambda\big]^L$. The key question in EVT is if $Q_L(\lambda)$, when $\lambda$ is appropriately shifted and scaled with respect to $L$, tends to a limiting distribution, i.e., \begin{equation}
	\lim_{L\rightarrow\infty}Q_L(a_L+b_Lz)=f(z)=:\exp\big[-g(z)\big],
	\end{equation}
	where $a_L$ and $b_L$ are size-dependent scale factors, and $f(z)$ and $g(z)$ are universal functions. The necessary condition for a size-independent limiting distribution is 
	\begin{equation}
	\lim_{L\rightarrow\infty}L\int_{a_L+b_Lz}^{\infty} \mathcal{P}(\lambda)d\lambda= \lim_{L\rightarrow\infty}L(a_L+b_Lz)^\text{-f}=g(z). 
	\end{equation}
	This gives $a_L=0$, $b_L=L^{1/\text{f}}$, and $g(z)=z^{-\text{f}}$. Therefore the distribution function $\mathcal{F}(z)=df(z)/dz$ of the scaled maximum eigenvalues $z=(L^{-1/\text{f}}\lambda_\text{max}-l)/s$ with location $l$ and scale $s$ is the Fréchet extreme value distribution 
	\begin{equation}\label{Frechet}
	\mathcal{F}(z)=\frac{\text{f}}{s}\frac{\exp\big[-z^{-\text{f}}\big]}{ z^{\text{f}+1}},
	\end{equation} 
	which establishes a direct relationship between the distributions of $\lambda$ and $\lambda_\text{max}$, both governed by the common exponent $\text{f}$. 
	
	As will be discussed in the following, since the average of the rescaled maximum eigenvalue $\tilde{\lambda}_{\text{max}}=L^{-\frac{1}{2}}\lambda_{\text{max}}$ is identified as the order parameter in the 2d Ising model, the scale factor $b_L$ can be inferred from the finite-size scaling relation  $\langle\tilde{\lambda}_{\text{max}}\rangle\sim L^{-\beta/\nu}$ with the known exact exponent $\beta/\nu=1/8$ as $b_L\sim L^{3/8}$ which indeed gives $1/\text{f}=3/8$. As shown by the solid line in the main Fig. \ref{fig2},  our proposed universal function (\ref{Frechet}) matches the data with parameters $l\simeq0.1$ and $s\simeq0.185$ related to the rescaled average and variance of our data, respectively.
	
 \textit{Critical mean-field ensembles.}
  To investigate the influence of the interaction range, we add $q-4$ randomly chosen {\it nonlocal} ferromagnetic interaction per spin, yielding an average coordination (number of neighbors) of $q$.  Rewiring is not allowed and any two spins cannot be directly connected by more than one link. 
		The long-range nature of the interactions effectively suppresses fluctuations and leads to a mean field limit of the Ising model. The self-consistency equation for the magnetization is 
	$\textsf{m}=\tanh\big[(q/T)\textsf{m}\big]$,
	which also provides the critical point $T_c=q$ for large $q$ (with interaction coupling and Boltzman constant set equal to $1$.)
	
\begin{figure}[t]
		\centering
		\includegraphics[width=3.3in]{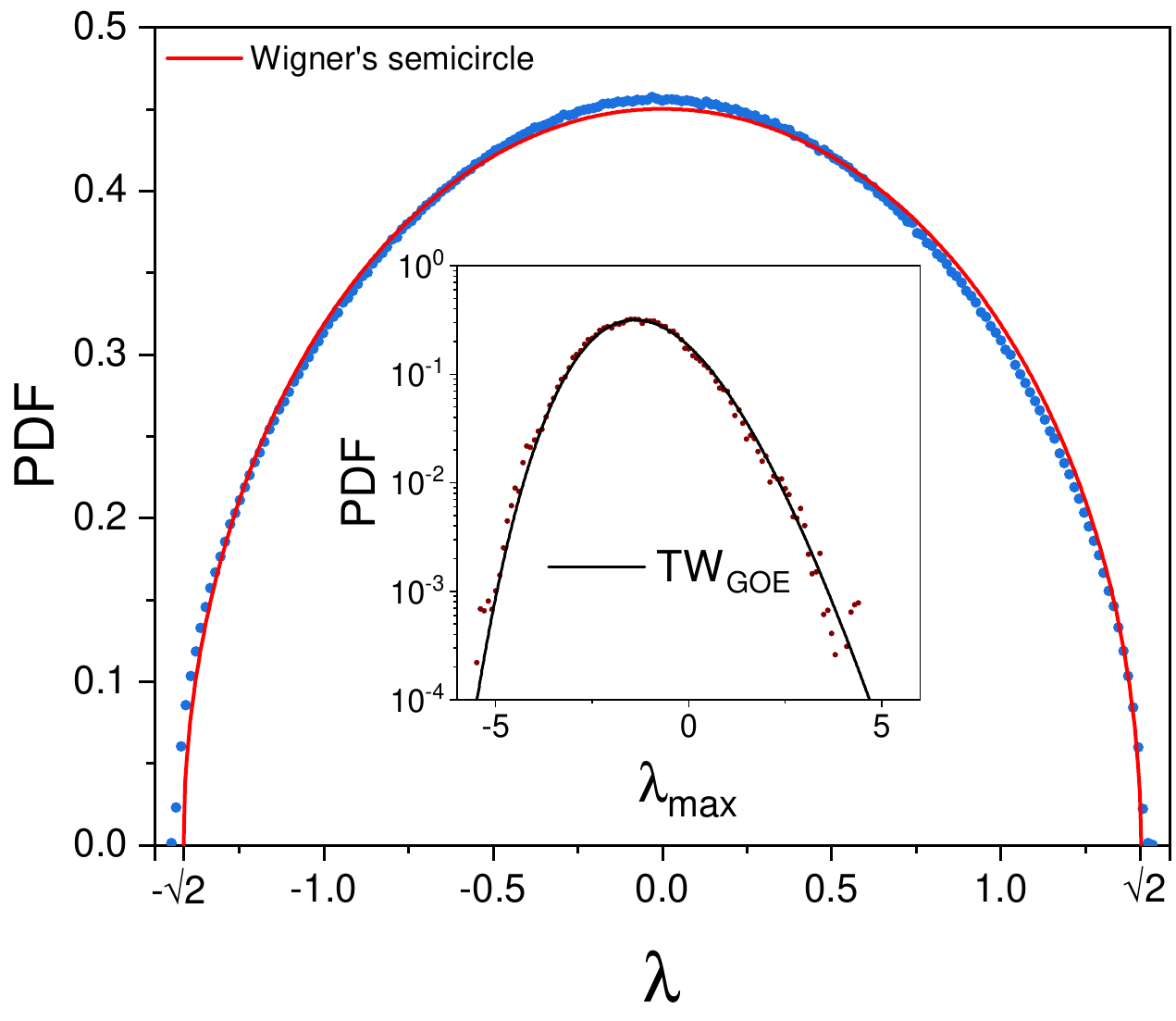}
		\caption{Main: Distribution of the eigenvalues of an ensemble of $10^5$ rescaled symmetric random matrices as a realization of a 2d Ising model with additional $10$ nonlocal links per spin ($q=14$) on a square lattice of linear size $L=2^{12}$, at its critical point $T_c=q$ (symbols) compared with the Wigner's semicircle law (the red solid line). Inset: Distribution of the maximum eigenvalues (symbols) sitting at the right edge $\lambda\simeq\sqrt{2}$ compared with the GOE $\mathcal{TW}_{1}$ distribution (the solid line).}
		\label{fig3}
	\end{figure}

	Figure \ref{fig3} shows the PDF of the eigenvalues for an ensemble of rescaled symmetric random matrices corresponding to models with $q=14,19$ and $32$ on a square lattice of size $L=2^{12}$, at the critical point $T_c=q$. Remarkably, the universality observed at the critical point of the 2D Ising model ($q=4$) is no longer present, having been replaced by Wigner's semicircle universality (main Fig. \ref{fig3}), with GOE $\mathcal{TW}_{1}$-distributed extremes (inset of Fig. \ref{fig3}). 
 
\textit{Off-Critical ensembles.}
The Ising model below the Curie point ($T<T_c$) exhibits nonzero spontaneous magnetization $\textsf{m}(T)$, which can be either positive or negative. 
This nonzero magnetization leads to an isolated extreme eigenvalue for each spin realization \cite{saber2022universal, malekan2022exact}, appearing as two disjoint symmetric Gaussian bumps around the semicircle-like spectrum of the bulk eigenvalues (see upper panel in Supplementary Figure S2). The second largest eigenvalues (in modulus) are located at the right edge of the semicircle, and their limiting distribution is known \cite{shcherbina2009edge} to follow the GOE $\mathcal{TW}_{1}$ distribution \cite{tracy1994level} (see lower panel in Supplementary Figure S2).  

Above the Curie point ($T>T_c$), thermal fluctuations destroy long-range correlations, causing the system to behave as if at the fixed point $T=\infty$ in the thermodynamic limit, where each spin takes the value $+1$ or $-1$ with probability $1/2$. With $\textsf{m}(T)=0$, the bumps vanish, and the eigenvalue distribution follows Wigner's semicircle law \cite{saber2022universal, malekan2022exact}. However, at finite $T$, due to finite-size effects and residual short-range correlations, we observe a deformed semicircle-like distribution for the bulk eigenvalues (see upper panel in Supplementary Figure S2), while the statistics of the extreme eigenvalues continue to follow the GOE Tracy-Widom distribution.

	\begin{figure}[t]
		\centering
		\includegraphics[width=3.4in]{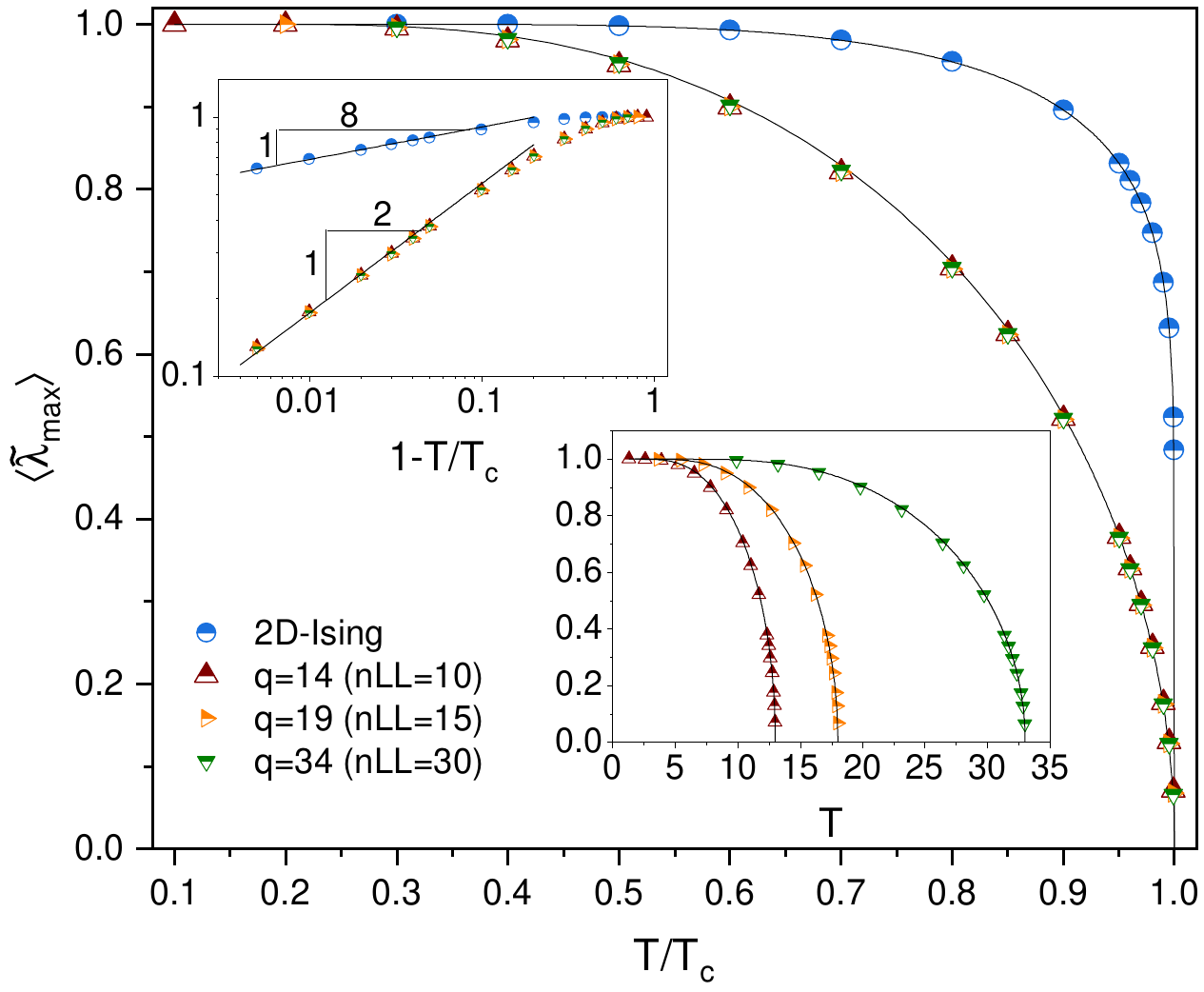}
		\caption{Main: The average rescaled maximum eigenvalue $\tilde{\lambda}_{\text{max}}=L^{-1/2}\lambda_{\text{max}}$ as a function of the reduced temperature $T/T_c$ for the pure 2d Ising model (half-filled circles) and the 2d Ising model with nonlocal interaction links with different $q=14, 19$ and $34$ (triangle symbols) compared with their corresponding exact solution for order parameter shown by the solid lines. Upper-left inset: Our data agree well with the scaling relation $\langle\tilde{\lambda}_{\text{max}}\rangle\sim (T-T_c)^\beta$ with the known exact exponents $\beta=1/8$ and $1/2$ for the pure 2d Ising model and the mean-field Ising model, respectively. Lower-right inset: $\langle\tilde{\lambda}_{\text{max}}\rangle$ versus temperature for different values of $q=14, 19$ and $34$ from left to right, respectively, that vanishes at $T_c=q$. }
		\label{fig4}
	\end{figure}

	\textit{Identification of interaction patterns.}
 Let us now consider the physical significance of the extreme eigenvalues. Based on our observation of the scaling relation $\langle\lambda_{\text{max}}\rangle\sim L^b$ with $b\simeq3/8$ at $T=T_c$ (Inset of Fig. \ref{fig2}), we confirm that the rescaled maximum eigenvalue $\tilde{\lambda}_{\text{max}}=L^{-\frac{1}{2}}\lambda_{\text{max}}$  plays the same role as the magnetization order parameter in the 2D Ising model. This rescaling supports the finite-size scaling relation  $\langle\tilde{\lambda}_{\text{max}}\rangle\sim L^{-\beta/\nu}$, with the known exact exponent $\beta/\nu=1/8$. Furthermore, we present the rescaled average maximum eigenvalue as a function of the reduced temperature $T/T_c$ in the main Figure \ref{fig4} (half-filled circles). We find excellent agreement between  $\langle\tilde{\lambda}_{\text{max}}\rangle$ and Onsager's magnetization for the 2D Ising model across all temperatures. Our data also supports the scaling relation $\langle\tilde{\lambda}_{\text{max}}\rangle\sim (T-T_c)^{\beta}$, with the known critical exponent $\beta=1/8$ for the order parameter in the 2D Ising model (see the upper-left inset of Figure \ref{fig4}). Additionally, we confirm the corresponding mean-field behavior in the presence of nonlocal interactions for different coordination numbers $q=14, 19$, and $34$ (see Fig. \ref{fig4}). The largest eigenvalue in the random matrix realization provides unique insights beyond standard bulk properties. It acts as an order parameter of the underlying physical systems, not only demonstrating a characteristic critical exponent but also revealing the number of nonlocal interaction links per site in the system. This connection is made explicit through the relationship
\begin{equation}
		q=\frac{T}{\langle\tilde{\lambda}_{\text{max}}\rangle}\tanh^{-1}\langle\tilde{\lambda}_{\text{max}}\rangle.
		\end{equation}
  The eigenvalue statistics in our matrix realization capture the critical fluctuations of the physical system, positioning random matrix analysis as a powerful tool for uncovering otherwise inaccessible properties of interacting systems. While the general properties of the magnetization distribution in the 2D Ising model at the critical point are well understood, its exact scaling function remains unknown in closed form. Our findings on the distribution of maximum eigenvalues may offer a pathway to derive the exact magnetization scaling function at $T=T_c$, potentially addressing a long-standing open problem in statistical physics.

\textit{Concluding remarks.}---
In this letter, we introduce a random matrix ensemble derived directly from a physical system, where the matrix elements correspond to the spins of a 2D Ising model at thermal equilibrium. By incorporating correlations from local interactions within the system, we study the spectral properties at the onset of criticality. We observe an emergent spectrum consistent with a $t-$distribution, featuring a power-law scaling tail characterized by an exponent $\text{f}=8/3$, which reflects the universal properties of the critical Ising model.  The largest eigenvalue $\lambda_\text{max}$ follows the Fréchet extreme value distribution with a shape parameter $1/\text{f}=3/8$, further supporting this universality. We find that the spectrum of the mean-field limit of the interacting model aligns with standard random matrix theory. Additionally, the average extreme eigenvalue acts as an order parameter, capturing both universal and non-universal aspects of the interaction structure in the underlying physical system.
	
\begin{acknowledgements}
 \textit{Acknowledgments.}---
 A.A.S. acknowledges partial support from the research council of the University of Tehran. We also thank the High-Performance Computing (HPC) Center at the University of Cologne, Germany, where a part of the computations was carried out. This work was in part supported by the Deutsche Forschungsgemeinschaft under grants SFB 1143 (project-id 247310070) and the cluster of Excellence 
ct.qmat (EXC 2147, project-id 390858490).
\end{acknowledgements}

	\bibliography{refs} 

\end{document}